\documentclass[copyright,creativecommons]{eptcs}
\providecommand{\event}{FOCLASA 2010} 
\usepackage{breakurl}             

\usepackage{amssymb}
\usepackage{amsmath}
\usepackage{latexsym}
\usepackage{theorem}
\usepackage{graphicx}
\usepackage{subfigure}
\usepackage{graphics}
\usepackage{epsfig}
\usepackage{makeidx}
\usepackage{showidx}

\usepackage{multicol}

\usepackage{color}
\usepackage{bcprules}

\usepackage{algorithm}
\usepackage{algorithmic}

\newcommand{\ie}{{\itshape i.e.}}
\newcommand{\eg}{{\itshape e.g.}}
\newcommand{\wrt}{{\itshape w.r.t. }}

\newcommand{\trans}[1][l]{\xrightarrow{\; #1 \;}}

\newtheorem{definition}{Definition}

\renewcommand{\algorithmiccomment}[1]{\textit{\footnotesize // #1}}

\title{Handling Data-Based Concurrency in Context-Aware Service Protocols
\thanks{This work is partially supported by the projects TIN2008-05932 and
P06-TIC-02250 funded by the Spanish Ministry of Science and
Innovation (MICINN) and FEDER, and the Andalusian local Government,
respectively.}}
\author{Javier Cubo
\institute{Dept. Computer Science\\ Univ. of M\'alaga, Spain}
\email{cubo@lcc.uma.es} \and Ernesto Pimentel \institute{Dept.
Computer Science\\ Univ. of M\'alaga, Spain}
\email{ernesto@lcc.uma.es} \and Gwen Sala\"un
\institute{INRIA-Grenoble\\ INP-LIG, France}
\email{Gwen.Salaun@inria.fr} \and Carlos Canal \institute{Dept.
Computer Science\\ Univ. of M\'alaga, Spain}
\email{canal@lcc.uma.es}}
\def\titlerunning{Handling Data-Based Concurrency in Context-Aware Service Protocols}
\def\authorrunning{J. Cubo, E. Pimentel, G. Sala\"un \& C. Canal}
\begin{document}
\maketitle

\begin{abstract}
Dependency analysis is a technique to identify and determine data
dependencies between service protocols. Protocols evolving
concurrently in the service composition need to impose an order in
their execution if there exist data dependencies. In this work, we
describe a model to formalise context-aware service protocols. We
also present a composition language to handle dynamically the
concurrent execution of protocols. This language addresses data
dependency issues among several protocols concurrently executed on
the same user device, using mechanisms based on data semantic
matching. Our approach aims at assisting the user in establishing
priorities between these dependencies, avoiding the occurrence of
deadlock situations. Nevertheless, this process is error-prone,
since it requires human intervention. Therefore, we also propose
verification techniques to automatically detect possible
inconsistencies specified by the user while building the data
dependency set. Our approach is supported by a prototype tool we
have implemented.
\end{abstract}

\section{Introduction}
\label{sec:introduction}

Service composition is a crucial paradigm in Service Oriented
Computing (SOC), since it allows to build systems as a composition
of pre-existing software entities, COTS ({\it
Commercial-Off-The-Shelf applications}) rather than programming
applications from scratch. An important issue of service composition
is to find out services with capabilities compatible to the user
requirements in order to compose them correctly.
In a traditional distributed environment, in which all the requests
are served in the same way, service composition is straightforward.
The introduction of Web-enabled hand-held devices has created the
necessity of a more context oriented composition in which the
produced response is aware of certain user and environment
information on the requesting client.
Thus, context-awareness enables a new class of applications in
mobile and pervasive computing, providing relevant information to
users. Therefore, context information can help users to find nearby
services, to decide the best service to use, to control reaction of
systems depending on certain situations, and so on.

Services are accessed through their public interfaces that may
distinguish four interoperability levels~\cite{Canal@LObjet06}: (i)
the {\it signature level} provides operation names, type of
arguments and return values, (ii) the {\it behavioural or protocol
level} specifies the order in which the service messages are
exchanged with its environment, (iii) the {\it service level} deals
with non-functional properties like temporal requirements,
resources, security, etc., and (iv) the {\it semantic level} is
concerned about service functional specifications (\ie, what the
service actually does). In industrial platforms, service interfaces
are usually specified using signatures (\eg,
WSDL\footnote{http://www.w3.org/TR/wsdl}), but some recent research
works~\cite{deAlfaroFSE2001,BBC@JSS05,cubo-coord07,YellinStrom97}
have extended interfaces with a behavioural description or protocol.
Protocols are essential because erroneous executions or deadlock
situations may occur if the designer does not take them into account
while composing clients and services~\cite{mateescu08,plasil02}. In
this way, service protocols evolving concurrently in a composition
need to impose an order in their execution if there exist data
dependencies. Dependency analysis is a technique to identify and
determine data dependencies between service protocols.
To the best of our knowledge, not many works have tackled the
handling of concurrent interactions of service protocols through
dependency
analysis~\cite{BasuEtAl-SCC07,CuboEtAl-CASTA09,Ensel-EDOC01,KuangEtAl-SCC07,Yan-JS08}.

In this work, we focus on systems that consist of
clients\footnote{In the sequel, we use client as general term
covering both client and user with a mobile device.} (users with a
mobile device such as a PDA or a smart phone) and services modelled
with interfaces constituted by context information, a signature, and
a protocol description (taking conditions into account). We also
consider a semantic representation of service instead of only a
syntactic one. We use OWL-S
ontologies\footnote{http://www.daml.org/services/owl-s/} to capture
the semantic description of services by means of relationships
between concepts within a specific domain.
In order to address the concurrency in the service composition in
these systems, we first formalise a model for context-aware clients
and service protocols. Second, we propose an approach to handle
dynamically the concurrent execution of context-aware service
protocols on the same user device, using mechanisms based on data
semantic matching. Our approach aims at assisting the user in
establishing priorities between these dependencies, avoiding the
occurrence of deadlock situations. Constraints on the concurrent
execution can be specified using a composition language which
defines operators for executing a sequence of protocols, a
non-deterministic choice between protocols, and for controlling the
data dependencies existing among several protocols executed at the
client level at the same time. In addition, since this process
requires human intervention (error-prone), we use analysis
techniques to automatically verify the correct execution order of
the protocols with respect to the built data dependency sets. Our
approach is supported by a prototype tool we have implemented. To
evaluate the benefits of our approach, we have applied it to
different case studies. We analyse the experimental results obtained
either with manual or interactive specification of data dependencies
and their corresponding execution priorities.

The rest of this paper is structured as follows.
Section~\ref{sec:model} presents our model formalising context-aware
clients and service protocols. In Section~\ref{sec:example}, we
introduce a case study we use throughout the paper for illustration
purposes. Section~\ref{sec:monitoring} presents the handling of
concurrent interactions of context-aware service protocols.
Section~\ref{sec:toolresults} describes the {\tt ConTexTive}
prototype tool that implements our approach, and shows some
experimental results. Section~\ref{sec:related} compares our
approach to related works. Finally, Section~\ref{sec:conclusions}
ends the paper with some concluding remarks.

\section{Context-Aware Service Model}
\label{sec:model}


\subsection{Interface Model} \label{sec:inteface}

Our model describes client and service interfaces using context
profiles, signatures and protocols. Context profiles define
information which may change according to client preferences and
service environment. Signatures correspond to operations profiles.
Protocols are represented using transition systems.

A context is defined as {\it ``the information that can be used to
characterise the situation of an entity. An entity is a person,
place, or object that is considered relevant to interaction between
a user and an application including the user and application
themselves''}~\cite{dey00}. Context information can be represented
in different ways and can be classified in four main
categories~\cite{kouadri03}: (i) user context: role, preferences,
language, calendar, social situation or privileges, (ii)
device/computing context: network connectivity, device capabilities
or server load, (iii) time context: current time, day, month or
year, and (iv) physical context: location, weather or temperature.
For our purpose, we only need a simple representation where contexts
are defined by context attributes with associated values. In
addition, we differentiate between static context attributes (\eg,
role, preferences, day, ...) and dynamic ones (\eg, network
connectivity, current time, location, privileges, ...). Dynamic
attributes can change continuously at run-time, so they have to be
dynamically evaluated during the service composition. Last, both
clients and services are characterised by public (\eg, weather,
temperature, ...) and private (\eg, personal data, bandwidth, ...)
context attributes. Thus, we represent the service context
information by using a {\it context profile}, which is a set of
tuples $(CA,CV,CK,CT)$, where: $CA$ is a context attribute or simply
context with its corresponding value $CV$, $CK$ determines if $CA$
is static or dynamic, and $CT$ indicates if $CA$ is public or
private (\eg, ({\it priv, Guest, dynamic, public}), where {\it priv}
is a public and dynamic context which corresponds to user privileges
with {\it Guest} as value).

A {\it signature} corresponds to a set of operation profiles. This
set is a disjoint union of provided and required operations. An
operation profile is the name of an operation, together with its
argument types (input/output parameters) and its return type.

A {\it protocol} is represented using a Labelled Transition System
(LTS) extended with value passing, context variables and conditions,
that we call Context-Aware Symbolic Transition System (CA-STS).
Conditions specify how applications should react (\eg, to context
changes). We take advantage of using ontologies to determine the
relationship among the different concepts that belong to a domain.
Let us introduce the notion of variable, expression, and label
required by our CA-STS protocol. We consider two kinds of {\it
variables}, those representing regular variables or static context
attributes, and variables corresponding to dynamic context
attributes (named context variables). In order to distinguish
between them, we will mark the context variables with the symbol
``$\sim$'' over the specific variable. An {\it expression} is
defined as a variable or a term constructed with a function symbol
$f$ (an identifier) applied to a sequence of expressions, $i \in
f(F_1,\ldots,F_n)$, $F_i$ being expressions.

\begin{definition}[CA-STS label]\label{def:ca-sts-label}
A {\it label} corresponding to a transition of a {\it CA-STS} is
either an internal action $\tau$ (tau) or a tuple $(B,M,D,F)$
representing an event, where: $B$ is a condition (represented by a
boolean expression), $M$ is the operation name, $D$ is the direction
of operations ({\it !} and {\it ?} represent emission and reception,
respectively), and $F$ is a list of expressions if the operation
corresponds to an emission, or a list of variables if the operation
is a reception.
\end{definition}


\begin{definition}[CA-STS Protocol]\label{def:ca-sts}
A {\it Context-Aware Symbolic Transition System} (CA-STS) Protocol
is a tuple $(A,S,I,Fc,T)$, where: $A$ is an alphabet which
corresponds to the set of CA-STS labels associated to transitions,
$S$ is a set of states, $I \in S$ is the initial state, $Fc
\subseteq S$ are correct final states (deadlock final states are not
considered), and $T \subseteq S \times A \times S$ is the transition
function whose elements $(s_1,a,s_2) \in T$ are usually denoted by
$s_1 \trans[a] s_2$.
\end{definition}

Finally, a {\it CA-STS interface} is constituted by a tuple
$(CP,SI,P)$, where: $CP$ is a context profile, and $SI$ is the
signature corresponding to a CA-STS protocol $P$.
Both clients and services consist of a set of interfaces. We assume
they have several protocols with their corresponding signatures, and
a context profile for each one. For instance, let us consider a
client with two different protocols $P_{c_1}$ and $P_{c_2}$. This
client consists of two interfaces such as: $I_{c_1} =
(CP_{c_1},SI_{c_1},P_{c_1})$ and $I_{c_2} =
(CP_{c_2},SI_{c_2},P_{c_2})$.


We adopt a synchronous and binary communication model (see
Section~\ref{sec:semantics-ca-sts} for more details). Clients can
execute several protocols simultaneously (concurrent interactions).
Client and service protocols can be instantiated several times.
At the user level, client and service interfaces can be specified
using: (i) context information into XML files for context profiles,
(ii) WSDL for signatures, and (iii) business processes defined in
industrial platforms, such as Abstract BPEL (ABPEL)~\cite{WSBPEL} or
WF workflows (AWF)~\cite{CSCPP-FACS07}, for protocols. Here, we
assume context information is inferred by means of the client
requests (HTTP header of SOAP messages), and we consider processes
(clients and services) implemented as business processes which
provide the WSDL and protocol descriptions.

\subsection{Operational Semantics of CA-STS} \label{sec:semantics-ca-sts}

We formalise first the operational semantics of one CA-STS service,
and second of $n$ CA-STS services. Next, we use a pair $\langle
s,E\rangle$ to represent an active state $s\in S$ and an environment
$E$. An environment is a set of pairs $\langle x,v\rangle$ where $x$
is a variable, and $v$ is the corresponding value of $x$ (it can be
also represented by $E(x)$). The function $type$ returns the type of
a variable. We use boolean expressions $b$ to describe CA-STS
conditions. Regular and context variables are evaluated in emissions
and receptions (by considering the current value of the context,
\eg, the current date), respectively. Therefore, two evaluation
functions are used in order to evaluate expressions into an
environment: (i) $ev$ evaluates regular variables or expressions,
and (ii) $ev_c$ evaluates context variables changing dynamically. We
define $ev$ as follows:

\begin{equation*}
ev(E, x) \triangleq
\begin{cases}
E(x) & \text{{\it if $x$ is a regular variable}}
\\
x &\text{{\it if $x$ is a context variable}}
\end{cases}
\end{equation*}

\begin{center}
$ev(E, f(v_1, \ldots, v_n)) \triangleq f(ev(E, v_1), \ldots, ev(E,
v_n) )$
\end{center}

Function $ev_c$ is defined in a similar way to $ev$, but it only
considers context variables. This is because we first apply $ev$ in
order to evaluate all the regular variables:

\begin{equation*}
ev_c(E, x) \triangleq E(x)
\end{equation*}

\noindent where $x$ is a context variable. We also define an
environment overloading operation ``$\oslash$'' such that, given an
environment $E$, $E \oslash \langle x,v \rangle$ denotes a new
environment, where the value corresponding to $x$ is $v$.

%
%

We present in Figure~\ref{fig:semantics-1} the semantics of one
CA-STS ($\trans[]_o$), with three rules that formalise the meaning
of each kind of CA-STS labels: internal actions $\tau$ (INT),
emissions (EM), and receptions (REC); and one rule to consider the
dynamic update of the environment according to the context changes
at run-time (DYN). Note that \wrt Definition~\ref{def:ca-sts-label},
$b \in B$ is a condition, $a \in M$ is an operation name, and $x \in
F$ and $v \in F$ correspond to a list of variables and expressions,
respectively. Condition $b$ may contain regular and/or context
variables and both of them must be evaluated in the environment of
the source service (sender), because the decision is taken in the
sender. However, evaluation of expressions $v$ only affects regular
variables (rule EM), since context variables will be evaluated in
the target service (receiver) to consider the context values when
the message is received (see rule COM in
Figure~\ref{fig:semantics-n}). We assume that the dynamic
modification of the environment will be determined by different
external elements depending on the type of context (\eg, user
intervention, location update by means of a GPS, time or temperature
update, and so on). Then, we model this situation by assuming a
transition relation which indicates the environment update, denoted
by $E \leadsto{_d} E'$, where $E'(x) \neq E(x)$ only if $x$ is a
dynamic context variable.

\begin{figure*}[!tbh]
\centering \small
\begin{minipage}[c]{0.45\linewidth}
\centering \infrule[INT]    {(s \trans[b,\tau] s') \in T \quad
                            ev_c(ev(E, b),b) = {\tt true}}
                            {\langle s,E\rangle \trans[\tau]_o \langle s',E\rangle}
\end{minipage}
\hfill \centering
\begin{minipage}[c]{0.45\linewidth}
\centering \infrule[REC]    {(s \trans[b,a?x] s') \in T \quad
                            ev_c(ev(E, b),b) = {\tt true}}
                            {\langle s,E\rangle \trans[a?x]_o \langle s',E\rangle}
\end{minipage}

\begin{minipage}[c]{0.6\linewidth}
\centering \infrule[EM]     {(s \trans[b,a!v] s') \in T \quad
                            ev_c(ev(E, b),b) = {\tt true} \quad v'=ev(E, v)}
                            {\langle s,E\rangle \trans[a!v']_o \langle s',E\rangle}
\end{minipage}
\hfill \centering
\begin{minipage}[c]{0.3\linewidth}
\centering \infrule[DYN]    {E \leadsto{_d} E'}
                            {\langle s,E\rangle \trans[\tau]_o \langle s,E'\rangle}
\end{minipage}

\caption{Operational Semantics of one CA-STS}
\label{fig:semantics-1}
\end{figure*}

The operational semantics of $n$ ($n > 1$) CA-STSs ($\trans[]_c$) is
formalised using two rules. A first synchronous communication rule
(COM, Figure~\ref{fig:semantics-n}) in which value-passing and
variable substitutions rely on a late binding
semantics~\cite{MilnerPW93} and where the environment $E$ is
updated. A second independent evolution rule (INE$_\tau$,
Figure~\ref{fig:semantics-n}). A list of pairs $\langle
s_i,E_i\rangle$ is represented by $[as_1,\ldots,as_n]$. Rule COM
uses the function $ev_c$ to evaluate dynamically in the receiver the
context changes related to the dynamic context attributes of the
sender. Regular variables have been evaluated previously in the rule
EM when the message is emitted. This dynamic evaluation handled in
the operational semantics allows to model service protocols
depending on context changes. Rule INE$_\tau$ is executed in case of
an internal service propagation that gives rise to either a state
(related to the rule INT) or an environment (rule DYN) change. Thus,
transitions $\trans[]_c$ do not distinguish between internal
evolutions coming from either internal actions in services or
dynamic updates in the environment.

\begin{figure*}[!tbh]
\centering \small
\infrule[COM]       {i,j \in
\{1..n\} \quad i \neq j \quad
                    \langle s_i,E_i\rangle \trans[a!v]_o \langle s'_i,E_i\rangle \quad \langle s_j,E_j\rangle \trans[a?x]_o \langle s'_j,E_j\rangle \\
                    type(x)=type(v) \quad E'_j=E_j \oslash \langle x,ev_c(E_j,v) \rangle}
                    {[as_1,\ldots,\langle s_i,E_i\rangle,\ldots,\langle s_j,E_j\rangle,\ldots,as_n] \trans[a!v]_c [as_1,\ldots,\langle
                    s'_i,E_i \rangle,\ldots,\langle s'_j,E'_j \rangle,\ldots,as_n]}

\medskip

\infrule[INE$_\tau$]{i \in \{1..n\} \quad
                    \langle s_i,E_i\rangle \trans[\tau]_o \langle s_i',E'_i\rangle}
                    {[as_1,\ldots,\langle s_i,E_i\rangle,\ldots,as_n] \trans[\tau]_c [as_1,\ldots,\langle
                    s'_i,E'_i \rangle,\ldots,as_n]}

\caption{Operational Semantics of $n$ CA-STSs}
\label{fig:semantics-n}
\end{figure*}

\section{Motivating Example}
\label{sec:example}

For illustration purposes, we consider a road info system that
consists of users travelling by car on a road and using mobile
devices (called Clients), and Info Services providing information
requested by the Clients.
Info Services contain information about routes, hotels, restaurants,
gas stations, multimedia entertainment such as movies, music,
images, shows, and so on, or museums. Some of these services are
free (\eg, Route or Gas Station Services) and others have to be
payed (\eg, Entertainment or Museum Services). For these latter ones
the Client needs to check his/her bank account.

For the sake of comprehension, we consider a reduced part of our
case study. Let us suppose that a Client, before starting the trip,
wants to plan a route. Afterwards he/she wants to perform at the
same time the purchase of both a music album to listen during the
trip, and a ticket for a museum located at his/her destination to
visit that same day. Ideally, the first request must be satisfied by
the nearest Route Service, which considers the context information
related to the Client location {\it loc} (dynamic context
attribute), and to the {\it traffic} and {\it weather} of the
environment (dynamic attributes). The nearest Entertainment Service
should manage the second request, by taking into account privileges
{\it priv} (dynamic attribute) of the Client (\eg, if the Client has
privileges of subscriber he/she will pay a reduced amount for an
album), and its {\it server load} (dynamic attribute). The third
request has to be replied by the Museum Service that also takes into
account Client privileges {\it priv}, and the {\it day} to visit the
museum (static attribute). This last request could also be replied
by the Entertainment Service, since this service can handle the
purchase of any show (museum, concert, cinema, etc) as well. We
consider all the context attributes mentioned are public and in our
scenario they have a default value, that in case of the dynamic ones
may change.

\begin{figure}[!tbh]
\centerline{\epsfig{figure=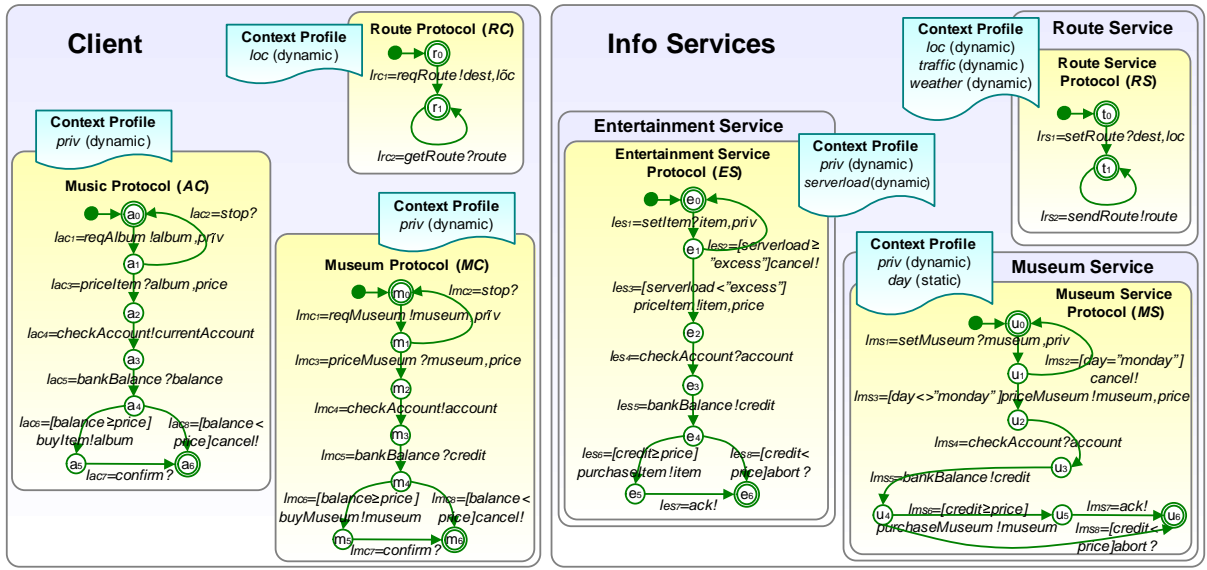,width=\linewidth}}
\caption[Case Study: CA-STS Protocols of Client and Info
Services]{CA-STS Protocols of Client and Info Services for our
Scenario} \label{fig:casestudy-protocols}
\end{figure}

This scenario requires to discover automatically the most
appropriate services for each client's request among the available
services (running at the moment of each request) from the repository
of Road Info Services. According to the dynamic nature of the
context information, context changes at run-time may occur. Thus,
for instance, once the Client has requested a route, when some
changes in his/her dynamic context attributes (\eg, {\it loc})
occur, the Route Service situated along the Client's way must
automatically recompute the route according to the new context
values (rule DYN, Figure~\ref{fig:semantics-1}). On the other hand,
we focus on the concurrent executions of several protocols at
run-time, that must be handled (\eg, the Client requesting
concurrently a music album and a museum ticket). All these
considerations make our approach appropriate to model this kind of
systems and to handle the concurrent interactions of protocols. For
the sake of simplicity, we suppose the available services from the
repository are Route, Entertainment and Museum Services. In
Figure~\ref{fig:casestudy-protocols}, the interfaces of Client and
Info Services are given for the scenario previously described.

The Client has three interfaces corresponding to the three client's
requests (Route, Music (Album) and Museum), which consist of three
protocols ($RC$, $AC$ and $MC$, respectively), each one with a
context profile, and a signature. The latter one will be left
implicit, yet it can be inferred from the typing of arguments (made
explicit here) in CA-STS labels. On the other hand, each service
(Route, Entertainment and Museum) has an interface with a context
profile, a (implicit) signature and a protocol ($RS$, $ES$ and $MS$,
respectively). We assume $RC$ should interact with $RS$, $AC$ with
$ES$, and $MC$ with $MS$. It is worth mentioning that the $ES$
protocol may be instantiated for communicating with different
client's requests related to movies, music, images, shows and so on.
Thus, $ES$ could also manage the Client's Museum request $MC$. Let
us consider, \eg, the label $RC:l_{rc_1}=reqRoute!dest,\tilde{loc}$
from the Client's Route protocol, where {\it dest} is a data term
which indicates the destination requested for the route, and {\it
$\tilde{loc}$} is a dynamic context attribute of the Client's Route
context profile. The Route Service protocol $RS$ receives the
request through a label such as $RS:l_{rs_1}=setRoute?dest,loc$
where {\it dest} and {\it loc} are variables.

\begin{figure}[!tbh]
\centerline{\epsfig{figure=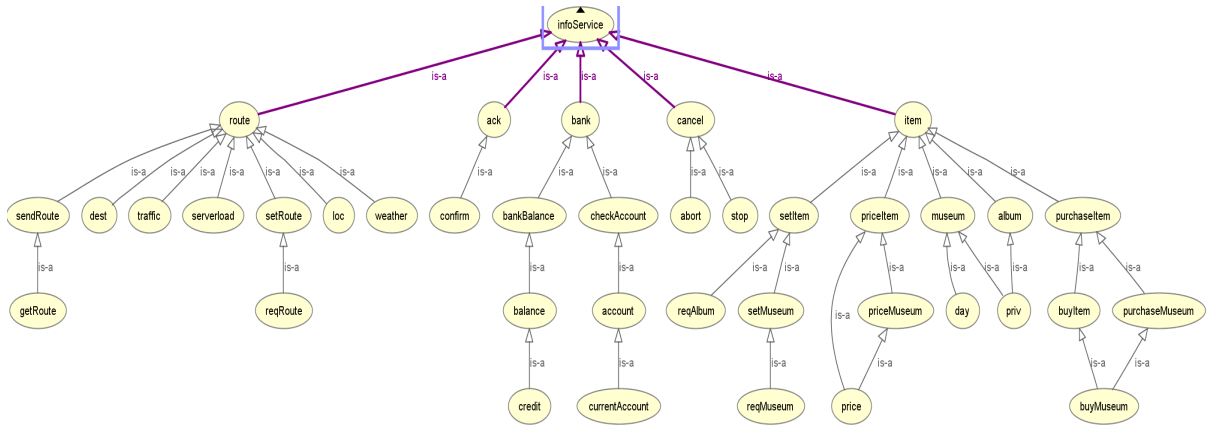,width=\linewidth}}
\caption[Road Info System Ontology]{Road Info System Ontology
generated using Prot\'eg\'e 4.0.2} \label{fig:roadinfo-ontology}
\end{figure}

Figure~\ref{fig:roadinfo-ontology} gives the domain ontology related
to this road info system. We present the classes used in our
scenario with their relationships. These classes represent concepts
which may be either a context attribute, an operation name, or an
argument. This ontology has been generated using Prot\'eg\'e
4.0.2\footnote{http://protege.stanford.edu/}.

\section{Handling Concurrent Interactions}
\label{sec:monitoring}

This section describes a composition language that allows to execute
and handle concurrently interactions between a client and several
services at the same time. Our language addresses data dependency
issues that appear in the concurrent execution of client protocols,
since all data received by a client are shared and can be accessed
by several of his/her protocols. Therefore, our mechanism allows to
maintain data consistency, even if a change occurs at run-time. We
can also detect problems coming from the data dependencies, which
would result in deadlocks during the execution of the protocols if
not corrected.

\subsection{Composition Language} \label{sec:language}

In this section, we formalise a language to dynamically compose
several protocols, with the following operators: {\it sequence},
{\it choice} and {\it parallel dependency} (or concurrency).

\subsubsection{Syntax} \label{sec:syntax}

A client can execute a {\it sequence} of the form $P_1 . P_2$, where
$P_1$ and $P_2$ are two protocols: ``execute $P_1$ and then $P_2$''.
A {\it non-deterministic choice} $P_1 + P_2$ can be performed: ``run
$P_1$ or $P_2$''. The concurrent execution of two protocols
$P_1,P_2$ is written $P_1||_{LD}P_2$: ``execute $P_1,P_2$ in
parallel while respecting data dependencies specified in $LD$''.
$LD$ is a set of {\it label dependencies} $\{(id:l>id':l')\}$, where
$l$ an $l'$ are labels, and $id$ and $id'$ are protocol identifiers
prefixing the labels. $LD$ represents dependencies between arguments
involved in the labels of these two protocols. Symbol ``$>$''
indicates the order of execution in which labels must be executed
(\eg, $(p_1:l>p_2:l')$, $l$ is executed before $l'$), being $l$ and
$l'$ the dominant and dominated labels, respectively. If more than
two protocols are executed concurrently, then we will detect the
data dependencies by pairs of protocols.
%
%
Here is the syntax of the composition language:

\begin{table}[!tbh]
\begin{center}
\begin{tabular}{llll}
  $P$ & ::= & $P_1 . P_2$ & {\it sequence} \\
    & $|$ & $P_1 + P_2$ & {\it non-deterministic choice} \\
    & $|$ & $P_1 ||_{LD} P_2$ & {\it parallel dependency}
\end{tabular}
\end{center}
\end{table}

The goal of our composition language is to illustrate with a minimal
expressiveness our service composition approach. We could have also
included for instance repetition operators such as $P\ast$ (executes
$P$ several times) or $P^x$ (executes $P$ $x$ times). Nevertheless,
repetition can be achieved by launching manually several times the
execution of $P$.

\subsubsection{Operational Semantics} \label{sec:semantics-language}

We formalise the operational semantics of the composition language.
The rules presented in Figure~\ref{fig:semantics-language} extend
the operational semantics of our model to the operators previously
considered. In SEQ2, $Fc_1$ refers to the correct final states of
the protocol $P_1$. Both $+$ and $||_{LD}$ are commutative,
therefore the symmetrical rules are omitted. Label $l$ represents
either the internal action $\tau$, an emission $a!v$, or a reception
$a?x$. PLD1 performs the concurrent execution of the protocols $P_1$
and $P_2$ \wrt a label dependency $(p_1:l>p_2:l')$, and removes the
label dependencies which include $l$ as first element from the label
dependency set $LD$. PLD2 works as PLD1, but without removing label
dependencies, since $l$ appears in a loop in its protocol. Last,
PLD3 executes a label which does not belong to the label dependency
set.

\begin{figure*}[!tbh]
\centering \small
\begin{minipage}[c]{0.3\linewidth}
\centering \infrule[SEQ1]   {\langle s_1,E_1\rangle \trans[l]_o
                  \langle s_1',E_1\rangle}
                 {\langle s_1,E_1\rangle . \langle s_2,E_2\rangle \trans[l]_o
                  \langle s_1',E_1\rangle . \langle s_2,E_2\rangle}
\end{minipage}
\hfill \centering
\begin{minipage}[c]{0.3\linewidth}
\centering \infrule[SEQ2]   {\langle s_2,E_2\rangle \trans[l]_o
                  \langle s_2',E_2\rangle \quad s_1 \in Fc_1}
                 {\langle s_1,E_1\rangle . \langle s_2,E_2\rangle \trans[l]_o
                  \langle s_2',E_2\rangle}
\end{minipage}
\hfill \centering
\begin{minipage}[c]{0.3\linewidth}
\centering \infrule[NDCH]   {\langle s_1,E_1\rangle \trans[l]_o
                  \langle s_1',E_1\rangle}
                 {\langle s_1,E_1\rangle + \langle s_2,E_2\rangle \trans[l]_o
                  \langle s_1',E_1\rangle}
\end{minipage}


\begin{minipage}[c]{0.45\linewidth}
\centering \infrule[PLD1]   {\langle s_1,E_1\rangle \trans[l]_o
                  \langle s_1',E_1\rangle \quad (p_1:l>p_2:l') \in LD
                  \quad \\ LD'=remove(l,LD) \quad \langle s_1,E_1\rangle \not \trans[l]_o\ast \langle s_1,E_1\rangle}
                 {\langle s_1,E_1\rangle ||_{LD} \langle s_2,E_2\rangle \trans[l]_o
                  \langle s_1',E_1\rangle ||_{LD'} \langle s_2,E_2\rangle}
\end{minipage}
\hfill \centering
\begin{minipage}[c]{0.45\linewidth}
\centering \infrule[PLD2]   {\langle s_1,E_1\rangle \trans[l]_o
                  \langle s_1',E_1\rangle \quad (p_1:l>p_2:l') \in LD \quad \\ \langle s_1,E_1\rangle \trans[l]_o\ast \langle s_1,E_1\rangle}
                 {\langle s_1,E_1\rangle ||_{LD} \langle s_2,E_2\rangle \trans[l]_o
                  \langle s_1',E_1\rangle ||_{LD} \langle s_2,E_2\rangle}
\end{minipage}

\infrule[PLD3]   {\langle s_1,E_1\rangle \trans[l]_o
                  \langle s_1',E_1\rangle \quad \forall ld \in LD (p_1:l \not \in get\_dominant\_label(ld) \wedge p_1:l \not \in get\_dominated\_label(ld))}
                 {\langle s_1,E_1\rangle ||_{LD} \langle s_2,E_2\rangle \trans[l]_o
                  \langle s_1',E_1\rangle ||_{LD} \langle s_2,E_2\rangle}

\caption{Operational Semantics of the Composition Language}
\label{fig:semantics-language}
\end{figure*}

We define formally the functions used in the operational semantics.
Function $remove(l,LD)$ eliminates the label dependencies which
include $l$ as first element from the label dependency set $LD =
\{ld_1,\ldots,ld_n\}$:
\begin{small} $remove(l,\{ld_1,\ldots,ld_n\}) = \{ld_i | ld_{i \in
\{1,\ldots,n\}} = (l_1 > l_2) \in \{ld_1,\ldots,ld_n\} \wedge l_1
\neq l\}$ \end{small}


\noindent Transition $\langle s_1,E_1\rangle \trans[l]_o\ast \langle
s_1,E_1\rangle$ is equivalent to $\langle s_1,E_1\rangle
\trans[]_o\ast \trans[l]_o \trans[]_o\ast \langle s_1,E_1\rangle$,
where $\trans[]_o\ast$ represents any sequence of transitions
$\trans[]_o$. This will determine if the label $l$ belongs to a loop
in transitions in a single protocol, starting from state $s$ and
ending in the same state $s$.
%
%
Functions $get\_dominant\_label$ and $get\_dominated\_label$ return
respectively the dominant and dominated labels from a label
dependency:

\begin{small} $get\_dominant\_label((id:l>id':l')) =
id:l; get\_dominated\_label((id:l>id':l')) = id':l'$ \end{small}

\medskip

Next, we describe two algorithms to detect label dependencies in
concurrent executions of protocols.

\subsection{Dependency Analysis} \label{sec:dependency}

Dependency analysis is a technique to identify and determine data
dependencies between service protocols. The main difficulty in
analysing dependencies for concurrent executions is how to obtain
the relationship between arguments.
Protocols evolving concurrently need to impose an order in their
execution if there exist data dependencies. A data dependency occurs
when a protocol receives a data, which is stored in the user device,
and when another client protocol accesses this data (\eg, wants to
send it). To detect and handle these dependencies, our
semi-automatic dependency analysis process consists of three steps:
(i) a first algorithm computes a set of pairs of label dependencies
between two protocols, (ii) the user makes a selection among these
pairs and determines the order of execution of the selected ones
(using the symbol ``$>$''), which allows to build an initial label
dependency set, and (iii) a second algorithm expands the
dependencies chosen by the user to a set as required by the semantic
rules PLD1, PLD2 and PLD3 formalised in
Figure~\ref{fig:semantics-language}.

The first step is performed by
Algorithm~\ref{algo:algo-pairs-label-dependencies}, that takes as
input two protocols, and a domain ontology. It returns all the label
dependencies among the argument types of the operation profiles of
both protocols. Our algorithm determines that two labels are
dependent by using the functions $degree\_match$, defined by
Paolucci {\it et al.}~\cite{paolucci02} (page 339), and $type$ to
compare their arguments and types, respectively. Function
$degree\_match$ defines four degrees of matching based on semantic
matching: {\tt \{exact,plugIn,subsume,\\fail\}}. The degree {\tt
fail} indicates that the two arguments compared do not match
semantically, so we do not consider that there exists a data
dependency between them. The remaining three indicate that there is
a semantic-based data dependency between the arguments.
\begin{algorithm}[!htb]
\scriptsize \caption{{\it
pairs\_label\_dependencies}}\label{algo:algo-pairs-label-dependencies}
\noindent \textit{returns a set of pairs of label dependencies for two protocols}\\
\noindent \textbf{inputs} protocols $P_1=(A_1,S_1,I_1,Fc_1,T_1)$ and $P_2=(A_2,S_2,I_2,Fc_2,T_2)$, ontology $Ont$\\
\noindent\textbf{output} a label dependency set $LD_p$\\
\begin{algorithmic}[1]
    \STATE $LD_p := \emptyset$ // initial value for set of pairs of label dependency
    \FORALL {$lp_1 \in A_1$}
        \STATE $A_{lp_1} := arguments(lp_1)$ // gets the arguments of $lp_1$
        \FORALL {$lp_2 \in A_2$}
            \STATE $A_{lp_2} := arguments(lp_2)$ // gets the arguments of $lp_2$
            \STATE $ATD := {\tt false}$ // by default no dependencies
            \FORALL {$arg_{lp_1} \in A_{lp_1}$}
                \FORALL {$arg_{lp_2} \in A_{lp_2}$}
                    \STATE $DM_{arg} := degree\_match(arg_{lp_1},arg_{lp_2},Ont)$
                    \STATE $DM_{typ} := type(arg_{lp_1}) = type(arg_{lp_2})$
                    \IF {$(DM_{arg} \neq {\tt fail}) \wedge DM_{typ}$}
                        \STATE $ATD := {\tt true}$ // argument and type dependency
                    \ENDIF
                \ENDFOR
            \ENDFOR
            \IF {$ATD$}
                \STATE $LD_p := LD_p \cup (p_1:lp_1,p_2:lp_2)$ // adds a pair
            \ENDIF
        \ENDFOR
    \ENDFOR
    \RETURN $LD_p$ // returns a set of pairs of label dependencies
\end{algorithmic}
\end{algorithm}
Function $arguments$ in
Algorithm~\ref{algo:algo-pairs-label-dependencies} returns all the
arguments belonging to a label $l = (b,m,d,f)$: \begin{small}
$arguments((b,m,d,f)) = f$ \end{small}


The complexity of Algorithm~\ref{algo:algo-pairs-label-dependencies}
is quadratic, O$(k \cdot (l \cdot a)^2)$, where $k$ is a constant
that indicates the number of dependent labels, and $l$ and $a$ are
the average numbers of elements in labels and arguments,
respectively.
In the second step, the set of pairs of label dependencies returned
by the previous algorithm is showed to the user. The user selects
the pairs of label dependencies he/she wants to preserve, and
chooses the execution order for each pair. The result is a label
dependency set. Given $LD_p = \{(p_1:l,p_2:l'),(p_1:l,p_2:l'')\}$,
if the user: (i) only selects the first pair appearing in $LD_p$,
\ie $(p_1:l,p_2:l')$, and (ii) indicates that $l'$ has to be
executed before $l$, then the result will be $LD =
\{(p_2:l'>p_1:l)\}$.

Last, Algorithm~\ref{algo:algo-extended-label-dependencies} takes as
input the two protocols of
Algorithm~\ref{algo:algo-pairs-label-dependencies} and the set
generated in the former step, and returns an extended label
dependency set.
%
%
The algorithm expands the set of label dependencies required by the
semantic rules PLD1, PLD2 and PLD3. For each label dependency $ld$,
the algorithm selects all the labels $pl_i$, $i \in \{1,\ldots,n\}$
preceding the dominant label of $ld$ in the corresponding protocol.
Then, for each $pl_i$ the algorithm adds a new label dependency
constituted by that $pl_i$ as dominant label and the dominated label
of $ld$ as dominated label. For instance, given two protocols $P_1$
with labels $l,l'$ in sequence and $P_2$ with $l''$, if $LD =
\{(p_1:l'>p_2:l'')\}$ is the label dependency set obtained in the
second step, then
Algorithm~\ref{algo:algo-extended-label-dependencies} returns a new
label dependency set $LD_e = \{(p_1:l>p_2:l''),(p_1:l'>p_2:l'')\}$.

\begin{algorithm}[!htb]
\scriptsize \caption{{\it
extended\_label\_dependencies}}\label{algo:algo-extended-label-dependencies}
\noindent \textit{returns an extended set of label dependencies from a label dependency set LD}\\
\noindent \textbf{inputs} protocols $P_1=(A_1,S_1,I_1,Fc_1,T_1)$ and $P_2=(A_2,S_2,I_2,Fc_2,T_2)$, label dependency set $LD$\\
\noindent\textbf{output} a label dependency set $LD_e$\\
\begin{algorithmic}[1]
    \STATE $LD_e := LD$ // sets the extended set equal to $LD$
    \FORALL {$ld \in LD$}
        \STATE $fl := get\_dominant\_label(ld)$ // gets the dominant label
        \STATE $sl := get\_dominated\_label(ld)$ // gets the dominated label
        \STATE $p_f := get\_id\_protocol(fl)$ // protocol $id$ of dominant label
        \STATE $p_s := get\_id\_protocol(sl)$ // protocol $id$ of dominated label
        \STATE $PL := get\_previous\_labels(fl,T_{p_f})$ // gets the previous labels to the dominant label $fl$ in the transitions $T_{p_f}$ of its protocol $p_f$
        \FORALL {$pl \in PL$}
            \IF {$(p_f:pl > ps:sl) \notin LD_e$}
                \STATE $LD_e := LD_e \cup (p_f:pl>p_s:sl)$ // adds a label dependency
            \ENDIF
        \ENDFOR
    \ENDFOR
    \RETURN $LD_e$ // returns the extended set of label dependencies
\end{algorithmic}
\end{algorithm}

Function $get\_id\_protocol$ gets the protocol identifier of a label
$(id:l)$: \begin{small} $get\_id\_protocol((id:l)) = id$ \end{small}


%

\noindent Function $get\_previous\_labels$ returns the labels
preceding a label $l$ in transitions $T$ of a protocol:

\begin{small} $get\_previous\_labels(l,T) = \{l' | \exists
(s_{i-1},l_i,s_i) \in T \wedge i = \{1,\ldots,n\} \wedge l_i = l'
\wedge l_{n+1} = l\}$ \end{small}


\medskip
%
%

The complexity of
Algorithm~\ref{algo:algo-extended-label-dependencies} is linear,
O$(ld \cdot TPL \cdot pl)$, where $ld$ is the number of label
dependencies, $TPL$ the average number of transitions to check in
the function $get\_previous\_labels$, and $pl$ the number of labels
preceding a concrete label.


{\bf Example.} Going back to our example, the Client wants to
execute the protocol $RC$ (route request) in sequence with the
parallel execution of the protocols $AC$ (music album request) and
$MC$ (museum ticket request): $RC.(AC||_{LD}MC)$. Our approach
builds the set of label dependencies between $AC$ and $MC$. First,
Algorithm~\ref{algo:algo-pairs-label-dependencies} takes as input
the two protocols, $AC$ and $MC$, and the domain ontology presented
in Figure~\ref{fig:roadinfo-ontology}. It returns a set of pairs of
label dependencies between $AC$ and $MC$: $LD_p =
\{(l_{ac_4},l_{mc_4}),(l_{ac_5},l_{mc_5}\}$ (protocol identifiers in
labels have been removed), since for the {\it checkAccount}
operation profile in both $AC$ and $MC$, {\it currentAccount} {\tt
exact} matches {\it account}, and for {\it bankBalance}, {\it
balance} is semantically compatible to {\it credit}, with degree of
match {\tt plugIn}. However, for instance, for {\it reqAlbum} and
{\it reqMuseum} of $AC$ and $MC$ respectively, the degree of match
of arguments and types is {\tt fail}, since although {\it
$\tilde{priv}$} {\tt exact} matches {\it $\tilde{priv}$}, {\it
album} {\tt fail} with respect to {\it museum}.
Then, the resulting set is given to the user, who selects the pairs
of label dependencies he/she wants to preserve, and chooses the
execution order for each pair. Let us suppose the user only selects
the pair $(l_{ac_4},l_{mc_4})$ to control the concurrent execution
of the operation $checkAccount$ in both $AC$ and $MC$, by executing
$l_{ac_4}$ before $l_{mc_4}$: $LD = \{(l_{ac_4}>l_{mc_4})\}$.
Last, Algorithm~\ref{algo:algo-extended-label-dependencies} takes
$LD$ as input and extends it with new dependencies needed to execute
the semantic rules PLD1, PLD2 and PLD3.
%
%
Thus, we obtain the final label dependency set: $LD_e =
\{(l_{ac_1}>l_{mc_4}),(l_{ac_2}>l_{mc_4}),(l_{ac_3}>l_{mc_4}),(l_{ac_4}>l_{mc_4})\}$.
This means that, \eg, for $(l_{ac_1}>l_{mc_4})$, $l_{ac_1}$ is
executed before $l_{mc_4}$, \ie, the label
$AC:l_{ac_1}=reqAlbum!album,\tilde{priv}$ is executed before the
label $MC:l_{mc_4}=checkAccount!account$, and so on. In such a way,
the algorithm controls that $l_{mc_4}$ will not be executed in $MC$
until $AC$ runs $l_{ac_4}$.

\subsection{Verification of Label Dependencies} \label{sec:verification}

The label dependencies construction process is error-prone, since it
requires human intervention. This process may provoke possible
inconsistencies which result in deadlocks during the execution of
the protocols according to the label dependency set computed
previously. Therefore, in this section we propose some verification
techniques to automatically detect these problems.

\begin{figure}[!tbh]
\centerline{\epsfig{figure=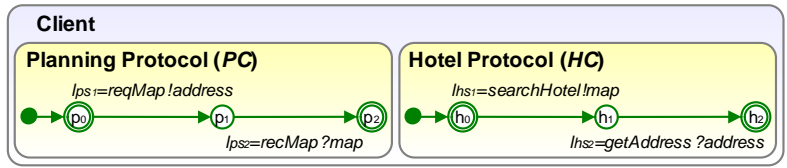,width=.65\linewidth}}
\caption[Protocol Concurrent Execution]{Client's Planning and Hotel
Protocols Executing Concurrently} \label{fig:verification-protocols}
\end{figure}

To illustrate the need of these verification techniques, we focus on
a simple example. In Figure~\ref{fig:verification-protocols}, we
give, {\it e.g.}, Client's Planning and Hotel protocols, $PC$ and
$HC$ respectively. The Planning protocol requests for a travel plan
to a specific address, and receives a map of the area close to that
address. The Hotel protocol searches for a hotel in that map, and
gets the destination address. By applying our dependency analysis,
Algorithm~\ref{algo:algo-pairs-label-dependencies} first returns the
pairs of label dependency: $LD_p =
\{(l_{ps_1},l_{hs_2}),(l_{ps_2},l_{hs_1})\}$. Second, let us suppose
the result of the user selection is: $LD =
\{(l_{hs_2}>l_{ps_1}),(l_{ps_2}>l_{hs_1})\}$. Last, the extended
label dependency set is: $LD_e =
\{(l_{hs_1}>l_{ps_1}),(l_{hs_2}>l_{ps_1}),(l_{ps_1}>l_{hs_1}),(l_{ps_2}>l_{hs_1})\}$.
In this set, the two label dependencies $(l_{hs_1}>l_{ps_1})$ and
$(l_{ps_1}>l_{hs_1})$ provoke a deadlock, since they are in mutual
exclusion. The two (crossed) label dependencies
$(l_{hs_2}>l_{ps_1})$ and $(l_{ps_2}>l_{hs_1})$ also generate a
deadlock, since the Planning protocol cannot start without the
$address$ and neither the Hotel protocol without the $map$.
%
%
The user will be informed to remove one of the label dependencies
which provoked this deadlock situation.

Algorithm~\ref{algo:algo-label-dependency-verification} takes as
input two protocols and their label dependency set, and returns a
set of traces (pairs of label dependency) leading to deadlock
states. To do that, the algorithm compares all the dominant labels
from the label dependency set with the dominated ones. If for two
label dependencies, a same label is dominant and dominated in both
directions or there exist crossed dependencies as described above,
then there is a deadlock situation. This problem has to be notified
to the user in order to let him/her modify the selection or
execution ordering of the label dependencies to avoid that
inconsistency.

\begin{algorithm}[!htb]
\scriptsize \caption{{\it
label\_dependency\_verification}}\label{algo:algo-label-dependency-verification}
\noindent \textit{detects possible inconsistencies specified in a label dependency set}\\
\noindent \textbf{inputs} protocols $P_1=(A_1,S_1,I_1,Fc_1,T_1)$ and $P_2=(A_2,S_2,I_2,Fc_2,T_2)$, label dependency set $LD$\\
\noindent\textbf{output} a deadlocked label dependency set $LD_d$\\
\begin{algorithmic}[1]
    \STATE $LD_d := \emptyset$ // initial value for set of pairs of deadlocked label dependency
    \FORALL {$ldp \in LD$}
        \STATE $fldp := get\_dominant\_label(ldp)$ // gets the dominant label
        \STATE $sldp := get\_dominated\_label(ldp)$ // gets the dominated label
        \STATE $p_{fldp} := get\_id\_protocol(fldp)$ // protocol $id$ of dominant label
        \FORALL {$ldg \in LD$}
            \IF {$ldp \neq ldg$}
                \STATE $fldg := get\_dominant\_label(ldg)$
                \STATE $sldg := get\_dominated\_label(ldg)$
                \STATE $p_{fldg} := get\_id\_protocol(fldg)$ // protocol $id$ of dominant label
            \ENDIF
            \IF {$((fldp == sldg) \wedge (sldp == fldg))$ \\ $\vee (sldg \in get\_previous\_labels(fldp,T_{p_{fldp}}) \wedge sldp \in get\_previous\_labels(fldg,T_{p_{fldg}}))$}
                \STATE $LD_d := LD_d \cup (ldp,ldg)$ // adds the pair of deadlocked label dependencies
            \ENDIF
        \ENDFOR
    \ENDFOR
    \RETURN $LD_d$ // returns the set of pairs of deadlocked label dependencies
\end{algorithmic}
\end{algorithm}

The complexity of
Algorithm~\ref{algo:algo-label-dependency-verification} is
quadratic, O$(ld^2 \cdot TPL)$, where $ld$ indicates the number of
label dependencies, and $TPL$ is the average number of transitions
to check in the function $get\_previous\_labels$.

{\bf Example.} For the scenario of our case study, we applied
Algorithm~\ref{algo:algo-label-dependency-verification} and checked
that no problems exist in the label dependency generated in
Section~\ref{sec:dependency}, since there is no trace (pair of label
dependency) that provokes a deadlock mismatch when executing
concurrently both protocols $AC$ and $MC$, \ie, $LD_d = \emptyset$.
Therefore, our label dependency set is correct.

\section{Tool Support and Experimental Results}
\label{sec:toolresults}


\subsection{Tool Support} \label{sec:tool}

\begin{figure}[!tbh]
\centerline{\epsfig{figure=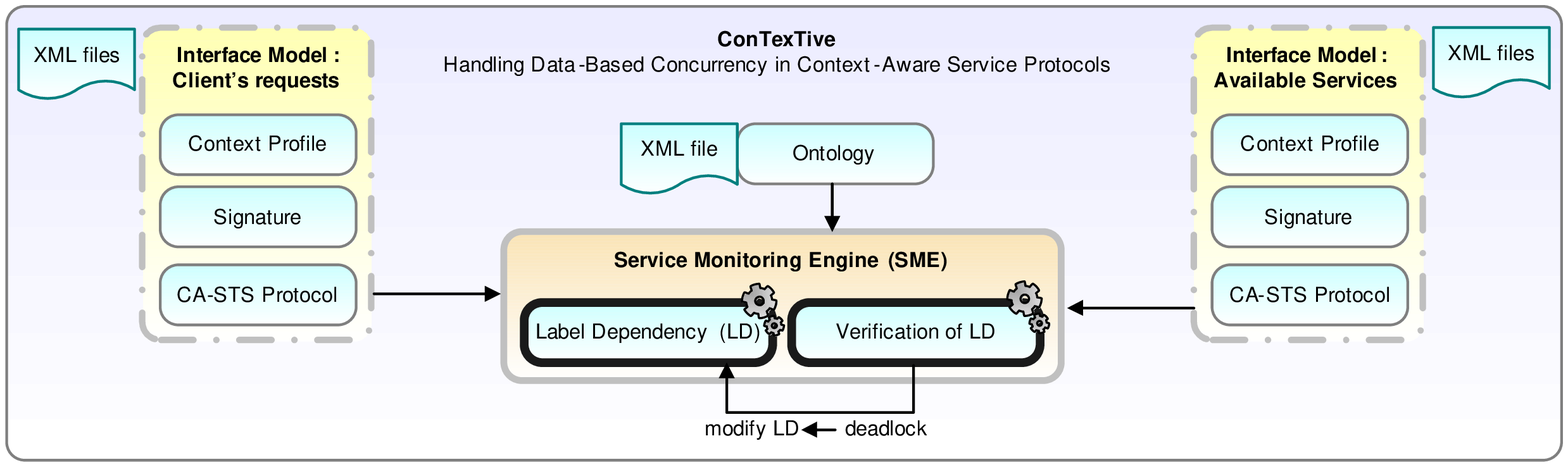,width=.75\linewidth}}
\caption[Tool Support Overview]{Tool Support Overview}
\label{fig:tool-overview}
\end{figure}

Our approach for handling concurrency of context-aware service
protocols, has been implemented as part of a prototype tool, called
{\tt ConTexTive}, which is integrated into our toolbox {\tt
ITACA}~\cite{ITACA-ICSE09}. {\tt ConTexTive} has been implemented in
{\tt Python} with the purpose of being incorporated inside a user
device. It aims at discovering services related to a client request
and handling the service composition by means of our composition
language. We have implemented the algorithms presented in this work,
in order to automatically detect data dependencies and check that
deadlock situations do not occur when executing protocols
concurrently.
Figure~\ref{fig:tool-overview} gives a tool support overview of how
our approach has been encoded. Our approach takes client and service
interfaces specified as XML CA-STSs, and an XML ontology of a
specific domain as input, and detects a label dependency set ($LD$)
for each pair of protocols executing concurrently, and checks if
this set is consistent (free of deadlocks).

{\tt ConTexTive} has been validated on several examples, such as an
on-line computer material store, a travel agency, an on-line booking
system or the case study presented here: a road info system.

\subsection{Experimental Results} \label{sec:results}

We have conducted a small experimental study with the assistance of
a group of volunteers. This study helped us to determine how our
approach behaves in terms of evaluating the benefits to find out
data dependencies in concurrent executions and to handle those
dependencies in terms of effort required, efficiency and accuracy of
the dependencies detected. Users performed tests either in a manual
or in an interactive (using the tool) way. In order to perform the
tests, we provided them a graphical representation of the interfaces
and a specific domain ontology to be used in the concurrent
interaction, for each problem. Each user solved different problems
using different specifications (manual or interactive\footnote{The
scenarios were executed on an Intel Pentium CPU 3GHz, 3GB RAM, with
Microsoft Windows XP Professional SP2.}) to prevent previous user
knowledge of a particular case study.
Table~\ref{tab:evaluation} shows the problems used for our study,
which are organised according to increasing size and complexity with
respect to the number of interfaces (client and services) involved
and the ontology, as well as the overall size of client protocols as
a total number of states and transitions. Tests considered all the
client protocols interacting concurrently. The table also includes
the comparison of experimental results using both manual and
interactive specification of data dependencies and their
corresponding execution priorities. We consider as parameters the
time required to solve the problem (in seconds), the number of label
dependencies detected (Depend. in Table~\ref{tab:evaluation}), and
the number of errors in the specified data dependency set.

\begin{table*}[!tbh]
\begin{center}
\begin{footnotesize}
\begin{tabular}{|c||c|c|c|c||c|c||c|c|c|c|}
\hline
&\multicolumn{4}{c||}{Size}&\multicolumn{6}{c|}{Parameter}\\
\cline{2-11}
Problem&\multicolumn{2}{c|}{Interfaces}&\multicolumn{2}{c||}{Client Protocols}&\multicolumn{2}{c||}{Time(s)}&\multicolumn{2}{c|}{Depend.}&\multicolumn{2}{c|}{Errors}\\
\cline{2-11}
&Client&Services&States&Transitions&M&I&M&I&M&I\\
\hline \hline
pc-store-v02&2&2&10&8&61,80&19,14&1&3&2&0\\
\hline
ebooking-v04&2&3&12&13&51,60&3,17&1&1&0&0\\
\hline
roadinfo-v06&3&3&16&18&113,62&16,51&5&4&3&0\\
\hline
travel-agency-v04&3&5&36&36&271,84&62,38&12&12&4&0\\
\hline
\end{tabular}
\end{footnotesize}
\caption{Experimental Results for the Manual (M) and Interactive (I)
Specifications} \label{tab:evaluation}
\end{center}
\end{table*}

As it can be observed in the results, there is a remarkable
difference in the amount of time required to solve the different
problems between manual and interactive specification. We measure as
errors the number of wrong, unnecessary or non-detected label
dependencies. Our tool always detects all the data dependencies and
it uses semantic matching to determine those dependencies, so this
is a clear advantage, which increases with the complexity of the
problem, compared to the manual specification. Thus, the time
elapsed for detecting dependencies by using our tool experiences a
linear growth with the size of the problem. Therefore, scalability
and performance of our tool are satisfactory, and in the worst case
(travel-agency-v04) the time required is roughly 1 minute, which is
a reasonable amount of time.

\section{Related Work}
\label{sec:related}

This section compares our approach to related works by emphasising
our main contributions. We successively describe works related to
service models by using protocols and/or context information, and
works focusing on monitoring of service composition to detect data
dependencies in the protocol interaction.


Context-based protocol models address the design and implementation
of applications which are able to be modified at the behavioural
interoperability level depending on context information. Not many
works have been dedicated to model context-aware service protocols.
Braione and Picco~\cite{braione04} have proposed a calculus to
specify contextual reactive systems separating the description of
behaviours and the definition of contexts in which some actions are
enabled or inhibited.
Related to context-aware adaptation, Autili {\it et
al.}~\cite{AutiliEtAl-FASE09} present an approach to context-aware
adaptive services. Services are implemented as adaptable components
by using the CHAMELEON framework~\cite{AutiliEtAl-Chameleon08}. This
approach considers context information at design time, but the
context changes at run-time are not evaluated.
In our approach, we propose a model to specify protocols based on
transition systems and extended with value passing, context
information and conditions, which has not been studied yet in
previous works. We consider context changes not only at design-time,
but also at run-time, since our model allows the continuous
evaluation of dynamic context attributes (according to the execution
of the operational semantics).


As regards concurrency, models for this discipline emerged, such as
CSP~\cite{hoare85} and CCS~\cite{milner80}, which address concurrent
systems from an algebraic perspective. The
$\pi$-calculus~\cite{milner99} builds on CCS as a process algebra
for communicating systems that allows expression of reconfigurable
mobile processes. Related to service concurrency, recent approaches
have been dedicated to the interaction of services at run-time with
the purpose of composing correctly their execution. In addition,
several works describe different ways to present data dependencies
according to their use for different purposes.
Vukovic~\cite{vukovic-cambridge07} presents an approach that focuses
on the recomposition of the composite service during its execution,
according to changes in the context. It provides a failure-tolerant
solution, but user preferences and control of independent requests
are not controlled, whereas our model supports that.
Mrissa {\it et al.}~\cite{Mrissa-ACM07} present a context-based
mediation approach to solve semantic heterogeneities between
composed Web services by using annotation of WSDL descriptions with
contextual details. Their architecture automatically generates and
invokes service mediators, so data heterogeneities between services
are handled during the composition using semantics and contexts.
These works do not handle data dependencies during the concurrent
execution of service protocols.
Basu {\it et al.}~\cite{BasuEtAl-SCC07} model such dependencies
using a directed edge between nodes. They generate a probabilistic
dependency graph as concatenation of all identified dependencies.
Ensel~\cite{Ensel-EDOC01} presents a methodology to automatically
generate service dependency model considering the direction of
dependencies.
In~\cite{KuangEtAl-SCC07}, Kuang {\it et al.} give a formal service
specification describing two types of dependency: dependency on
assignment (between the input and output interfaces of an
operation), and dependency on sequence (order among operations of a
service).
Yan {\it et al.}~\cite{Yan-JS08}, propose an approach to discover
operation dependencies using semantic matching of input and output
interfaces and the invoking order among operations. They construct
frequency and dependency tables in order to derive indirect
dependency relationships by transitive closure algorithm.
Most of these approaches do not consider a combination of both the
directionality and the execution order to detect dependencies. To
the best of our knowledge, the only attempt taking both restrictions
into consideration is~\cite{Yan-JS08}. Compared to these related
works, our approach does not only detect data dependencies,
addressing both direction and order, that appeared between
communications, but it also allows context-aware protocol concurrent
executions at run-time by means of our composition language. In
addition, in order to analyse dependencies we rely on semantic
matching techniques between data.

\section{Concluding Remarks}
\label{sec:conclusions}

In this paper, we have described a model to formalise context-aware
clients and services. We have also proposed a composition language
to handle dynamically the concurrent execution of service protocols.
We have defined algorithms to detect data dependencies among several
protocols executed on the same user device. These algorithms make
possible to establish some priorities on the concurrent execution of
protocols affected by these dependencies. In this way, our approach
allows to maintain data consistency, even if a parallel change
occurs at run-time. Last, we have proposed verification techniques
to automatically detect possible inconsistencies specified by the
user while building the data dependency set.
We have implemented a prototype tool, {\tt ConTexTive}, which aims
at handling the concurrent interaction of service protocols at
run-time. Our approach focus on mobile and pervasive systems.

We are currently working on avoiding the human intervention in the
process of building the data dependency set by means of priorities
previously defined. This will allow to determine automatically the
execution order of the detected dependencies, reducing the time
required in the interactive specification. We are also extending our
approach to solve other problems arisen in the context-aware service
composition, such as exception or connection loss.
As regards future work, our main goal is to incorporate our
prototype tool inside a user device in order to support concurrency
in real-world applications running on mobile devices. We also plan
to extend our framework to tackle dynamic reconfiguration of
services, handling the addition or elimination of both services and
context information. In addition, we want to include the repetition
operators in our composition language, and to handle concurrent
execution of more than two protocols by detecting at the same time
all data dependencies existing among several protocols.

\bibliographystyle{eptcs} 
\begin{small}
\bibliography{biblio,biblio2,biblio3,biblio4,biblio5,biblio6}
\end{small}
%
%
\end{document}